\documentstyle[preprint,aps]{revtex}

\begin{document}

\draft

\title{Curvature and Chaos in General Relativity}

\author{Werner M. Vieira\thanks{e-mail: vieira@ime.unicamp.br} and
 Patricio S. Letelier\thanks{e-mail: letelier@ime.unicamp.br}}
\address{Departamento de Matem\'atica Aplicada\\
Instituto de Matem\'atica, Estat\'{\i}stica e Ci\^encias da
 Computa\c{c}\~ao\\
Universidade Estadual de Campinas, CP 6065\\13083-970
 Campinas, SP, Brazil
\\{\rm August 7, 1996 }}

\maketitle

\begin{abstract}

We clarify some points about the systems
 considered by Sota, Suzuki and Maeda
in Class. Quantum
 Grav. {\bf{13}}, 1241 (1996).
 Contrary to the authors' claim
for a non-homoclinic kind of chaos, 
 we show the chaotic cases
 they considered are
 homoclinic in origin. 
The power of local criteria to predict
 chaos is once more questioned.
We find that their local, curvature--based
 criterion is
 neither necessary nor sufficient for the occurrence of chaos.
 In fact, we argue that a merit of their search
 for local criteria applied to General Relativity
 is just to stress
 the weakness of locality itself, free of any pathologies 
 related to the motion in effective Riemannian geometries.

\end{abstract}


\section{Introduction}

In a recent paper Sota, Suzuki and Maeda~\cite{Sota}
found new support to the  important issue of chaotic
behavior in General 
Relativity (GR)~\cite{HobillBurdColey,WernerPatricio}.
 Based on 
the sign structure of the eingenvalues of Weyl's tensor,
they also
proposed what should be a
 sufficient, local criterion for the
occurrence of chaos in GR 
(we will refer to this criterion as SSM after the authors).
 They restrict the analysis only 
to the case of geodesic motion of test particles in fixed, 
exact spacetime geometries, more specifically, in static 
axisymmetric spacetimes in vacuum.

In section 2  firstly we clarify some misunderstandings about
 the systems
 considered in~\cite{Sota} 
related to the SSM statement. Contrary to the
 authors' claim, we show that the 
chaotic cases presented in~\cite{Sota} have a homoclinic origin.
 We also emphasize that
 the families of unstable periodic orbits (UPOs) related to the
 homoclinic tangle are not the same in the static cases
considered in~\cite{Sota}
 and in the time--depending
cases studied for example in~\cite{BombelliCalzetta}.

In section 3 we find that SSM
 is neither necessary nor
 sufficient as a criterion to predict chaos.
We address this insufficiency in relation to the deeper question
 whether local criteria can predict chaos. The
 central idea and the numerical results of~\cite{Sota} become 
important to this end. As space--time is intrinsically
 Riemannian in GR,
 we argue that a main merit of Sota, Suzuki and
 Maeda's search
 for local criteria applied to GR does not
 have yet been noted and it is just to exhibit
 the weakness of locality itself, free of any pathologies
related to the motion in effective Riemannian geometries.

\section{The underlying homoclinic dynamics}

We show in this section that the idea of existing two types of
 chaos in the systems studied in~\cite{Sota} is wrong.
 The first type should be 
associated to
 the so called homoclinic tangle of the unstable/stable manifolds
 departing/arriving from/at an UPO. It
 should rise due to the breaking of the reflection
 symmetry of the system about the middle plane (as seen in fig.\ 10
 of ~\cite{Sota}). The
 second ``new'' kind of chaos should rise when that symmetry
 is preserved
 (as seen, e.g., in fig.\ 4 of~\cite{Sota})
 and should be explained
by the occurrence of certain locally unstable regions
 (named LU regions in that paper) rather than a homoclinic tangle. 
We show in the following that, also in the
 latter case, the evidences for the homoclinic tangle is a matter
 of figure scale allied to a careful
 search of the structures involved in the tangle.
 It seems that this was what
 improperly motivated the 
statement of SSM as a sufficient criterion.

Consider the 2--Curzon system with the same numeric
 values for energy, angular momentum, etc.\ of fig.\ 4.
 This configuration exhibits chaos yet preserving the 
reflection symmetry about the middle plane. We 
found that there exists an UPO {\it{confined}} to the plane
 $z, p^z$ and characterized by $p^\rho =0$ and
 $\rho=\rho_0 \approx 4.4833 R_0$ where $R_0=GM/c^2$
 (this is a local maximum of the
 effective potential in the middle plane).
 This UPO is transversal to the middle plane
$z=0$. On the other hand, the  geodesic flow
has reflection symmetry
 about that plane in the four--dimensional phase space
 ($\rho,z,p^\rho,p^z$).
 Both facts makes the middle plane
 ideal, as Poincar\'e's section, to unravel the whole homoclinic
 tangle: the UPO itself will appear in the section
 as a fixed point and 
the four (two--dimensional) unstable/stable
manifolds emanating from the UPO
 will intersect the section in four (one--dimensional)
 X--type branches 
departing/arriving from/at the fixed 
point (at the center of the X).

We stress that the (one--dimensional)
 UPO above plus its associated (two--dimensional) manifolds 
 must not be confused with an
unstable equilibrium {\it{point}} plus the (one--dimensional)
 unstable/stable manifolds
 emanating from it.
 In the case with reflection
 symmetry, the last structure does also exist and
 lies confined to the middle plane.
 Although immersed in the same 
four--dimensional phase space,
both structures are in different energy surfaces and hence do not 
share with the same Poincar\'e's section. We also mention that
 by a suitable
periodic perturbation we can break the
 integrability
 around the unstable equilibrium point yet maintaining the
 dynamics confined to the 
middle plane, as studied, e.g.,
 in~\cite{BombelliCalzetta}.
In the latter case we are left with a so called
 one and a half degree of freedom system
 in the extended three--dimensional phase
 space ($\rho,p^\rho,t$).
 In this case
 the Poincar\'e's sections are better constructed
 as discretized maps of the continuos flow in the plane
($\rho,p^\rho$) at times $t=n T, n=0,1,2,\cdots,$
 where $T$ is the period of the perturbation.
In the present static situation with reflection
 symmetry we deal with the whole four--dimensional phase space yet 
preserving the integrability of the 
family of orbits lying in the 
middle plane, to which the unstable 
equilibrium point together with its homoclinic orbit pertains.
 
Indeed, all that is shown in our figure 1. Fig.\ 1--a is basically
 the same as fig.\ 4 of~\cite{Sota} except that we add the elements
 related to the UPO namely, the central fixed
 point and its X--type branches (only the two
 right branches are shown in fig.\ 1--a), and
 also the boundary of the motion, which is itself one among the whole
 family of closed orbits confined to the middle plane. If we amplify
 the left region of fig.\ 1--a and specialize the study there, we
 obtain fig.\ 1--b. We see that among the four branches associated
 to the central fixed point, only the two on the left are chaotic 
(at least in the figure scale). For the sake of clarity we show
 in fig.\ 1--b only a half of the full homoclinic 
tangle, i.\ e.\ , only that associated to the unstable left
 branch. It was numerically obtained by integrating about 1000 
starting points equally spaced on the segment $\rho/R_0 \in
 [4.427,4.477]$, on the straight line given by 
equation $p^\rho /\mu c = 0.00461478 \times (\rho/R_0-4.477)$. The
numerical and hence approximate character of this search 
is responsible for the filling of the inner region of the
 half--tangle. The overall figure has about 19000 points. In
 other words, geodesic equations are a dynamical system 
prior of  being 
 a relativistic one. As such, it must exhibit the 
universal features --- the homoclinic tangle among
 them --- expected
 within the general context of dynamical systems.

We note
that all this remains valid for any other figure 
of~\cite{Sota}. The question is that finding UPOs
 in phase space
is in general
 a very difficult task. Moreover, given an UPO,
 the stable/unstable manifolds emanating from it have a
 rather complicated cylindrical topology in
 the full phase space~\cite{Werner}. Then, in generic cases,
we need a correspondingly 
complicated {\it{nonplanar}} Poincar\'e's section
to follow and suitably intersect those cylindrical manifolds 
to see the related homoclinic figure. 
Cases like
 our fig.\ 1 are possible only when
 the symmetries of the motion are strong enough to
constrain the cylindrical manifolds emanating from the UPO 
to remain
 always longitudinal to a planar section. In this sense,
fig.\ 1 provided by the 2--Curzon system is an amazing
exact relativistic example of full homoclinic
 figure captured from a differential system.

\section{Curvature, Chaos and General Relativity}

The most celebrated conclusion (and the only really sound 
until now) about the relation between local, curvature--based
 properties and chaoticity of the  motion is the following:
 the geodesic flow on a compact manifold with all sectional
 curvatures negative at every point is
 chaotic~\cite{ArnoldAvez}. This very special case 
has inspired the search for related extensions in 
both Newtonian theory and GR. In the Newtonian theory,
 we apply Maupertuis' principle
in the context of Hamiltonian
 dynamics to provide the motion with an effective 
Riemannian manifold and search for
 criteria able to predict chaotic behavior 
from its
 curvature properties
 (see for instance~\cite{Arnold,Szy7}). The application of 
this method to cosmological models~\cite{Szy8,Szy5}
intends mainly to formulate gauge invariant criteria for the 
occurrence of chaos in GR.

An alternative to assure gauge invariance to the
methods lying on Maupertuis' principle is proposed in~\cite{Szy2}
and applied with little changes
to the Bianchi IX cosmological model~\cite{Biesiada}.
The method consists also of
 local analysis of
sign structure of eingenvalues, this time
based on invariant curvature polynomials constructed after the
 conformal transformation of the metric
(this contrasts, for instance, with 
Lyapunov's exponents approach, 
which is a global yet highly gauge dependent method).
 The method points to the occurrence
of local instability in
the Bianchi IX model near the singularity. This adds to a
plethora of early analysis and numerics
 (see for instance the references cited in~\cite{Biesiada}), which 
 reinforces even more the increasingly accepted idea that
 Bianchi IX is after all chaotic in
some meaningful sense.

This line of 
search, yet interesting, remains inconclusive 
to the end of predicting chaos, for several reasons: 
i) {\it{Locality}}. The method deals with the geodesic deviation 
equation on the effective manifold, however, local instability of
 the geodesic flow does not imply chaos, even when the 
curvature is everywhere negative, as exemplified 
in~\cite{Yurtsever}. ii) {\it{Averaging}}. It ever involves 
some kind of average over the geodesic deviation equation
  to produce quantitative information about chaos, which
 may suppress important aspects of the true motion and
 even lead to the lack of gauge invariance of the quantities and 
iii) {\it{Pathologies}}. The effective manifold exhibits 
boundaries which can trouble the understanding of the
 true motion in the physical manifold. These problems
 are pointed out in~\cite{Szy7} and discussed in details
 in~\cite{BurdTavakol,Szy5}; we also mention that 
in a recent paper Szyd\l owski
et.\ al.~\cite{Szy1} propose to circumvent the last difficulty
above at
 the price of abandoning smooth manifolds in favor of lesser
stringent differential spaces.

The good idea behind~\cite{Sota} is to work directly in
 spacetime in GR,
 since it is intrinsically Riemannian, and to retain a
 truly local approach
of the geodesic deviation equation. This procedure has 
the merit of testing  locality itself, free of any artifacts
involving mean motion or boundaries of an effective
 manifold (in a second part of their paper the authors
 make averages for some quantitative estimates, which 
is of no concern here). So, possible relations between 
curvature and chaos can be studied directly in the 
physical configuration space, leading eventually
 to more clean conclusions about them.

SSM proposed
 in~\cite{Sota} is just an example of such a guessed 
relation: what is the role of the sign structure of the 
eingenvalues
 of the curvature tensor for the onset of chaos? They 
find correlations that, although interesting, are not 
enough to state a criterion
 for the occurrence of chaos in GR. In fact,
 these correlations associated to the existence of
 certain locally unstable regions (LU regions),
 are neither necessary nor sufficient to predict chaos,
 as it is
already clear from the numerical study made by themselves.
 This 
can be clearly seen for instance in figures 4 and 10 of their
 paper. In
 fig.\ 4  there exists a region of preserved (small) tori,
 surrounded by the larger chaotic region all being totally
 immersed
 in a LU region, while in fig.\ 10 the LU region does not meet
 anywhere the chaotic region. Probably, the authors' claim for
 the sufficiency of SSM had its motivation in the subtleties 
already discussed in section 2. This just exhibits
 the weakness of locality itself, free of any problems
related to the motion in effective Riemannian geometries.

 As expected, we can conclude that any local analysis, in
 effective or even physical spaces, is far from being 
sufficient
to predict a global phenomenon like chaotic motion.
 Conversely however, it is natural to suppose
 that chaoticity, if it is there, exerts influence on certain local
 properties, e.g., on the sign structure of the eingenvalues
of the  curvature tensor in that neighborhood.
 Obviously, this in no way prevent us to find partial methods 
that may in some sense feel the presence of chaos. 
This is all we need in many practical situations.
However, they are not sufficient for the study of the
 role played by chaos as a basic phenomenon. In this
 direction, it is clear that we need more elaborated 
prescriptions to approach this rather complex
 subject; as dealt with, for instance, in~\cite{Yurtsever}.

The authors thank 
 CNPq and FAPESP for financial support.

\begin{figure}

\caption{Poincar\'e's sections through the plane $z=0$ in Weyl
 coordinates for timelike geodesic orbits in 2--Curzon geometry, with 
nondimensional parameters
$E^2=0.913 (\mu c^2)^2$ and $L=6.94 G\mu M/c$, where $M$ is the (same)
 mass of the singularities located at $\pm 2GM/c^2$ on the z axis
 and $\mu$ is the mass of the test particle. In (a) we show the
 main global structures of the section namely, the small (on the
left) and the great (on the right) regions of preserved tori
 separated by the X--type unstable/stable branches centered at
the fixed point corresponding to the UPO (only the two right
 branches are exhibited here) and the closed boundary orbit lying 
in the middle plane.
 Part (b) is an
 amplification of the left region of (a) and shows 
a half of the homoclinic tangle surrounding the small tori,
 associated to the 
left (unstable) branch departing from the fixed point,
 all immersed in a LU region.}\label{fig1}

\end{figure}


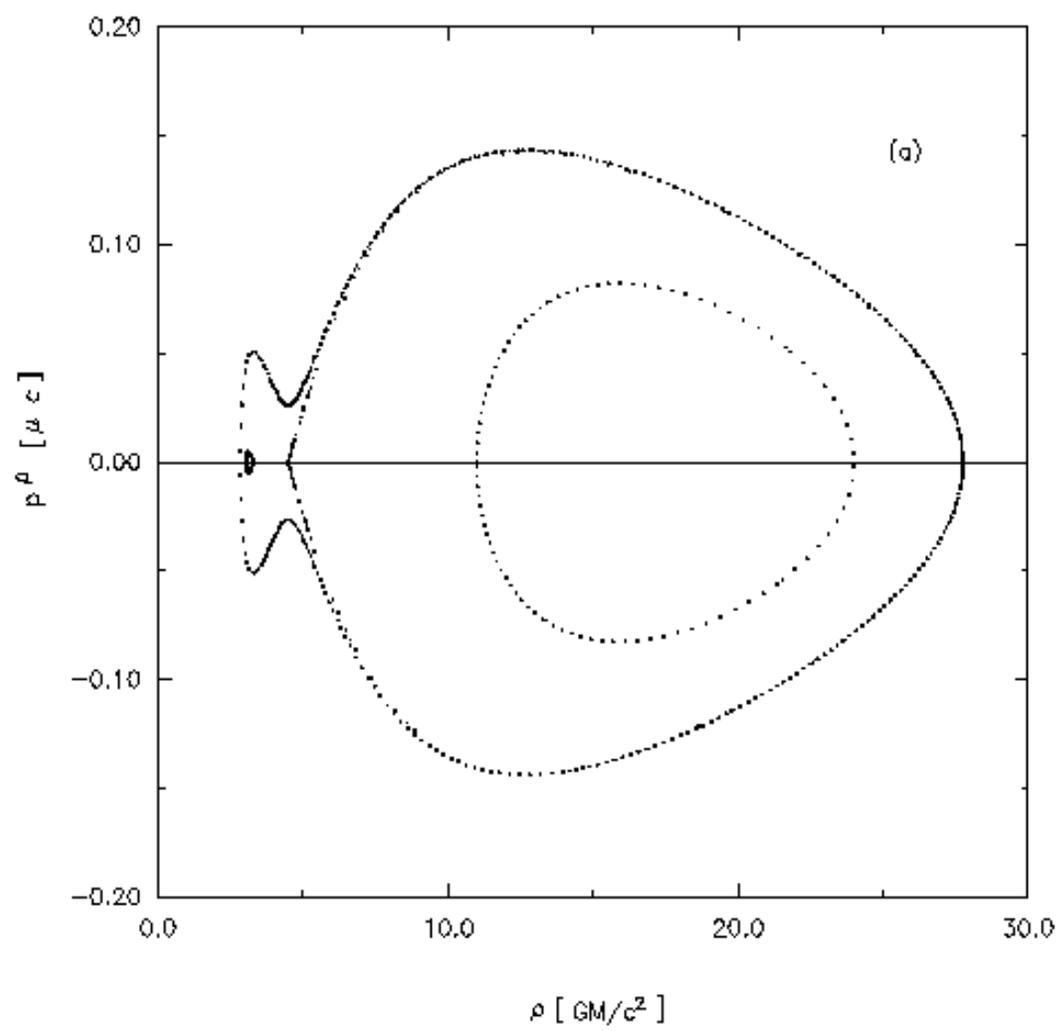
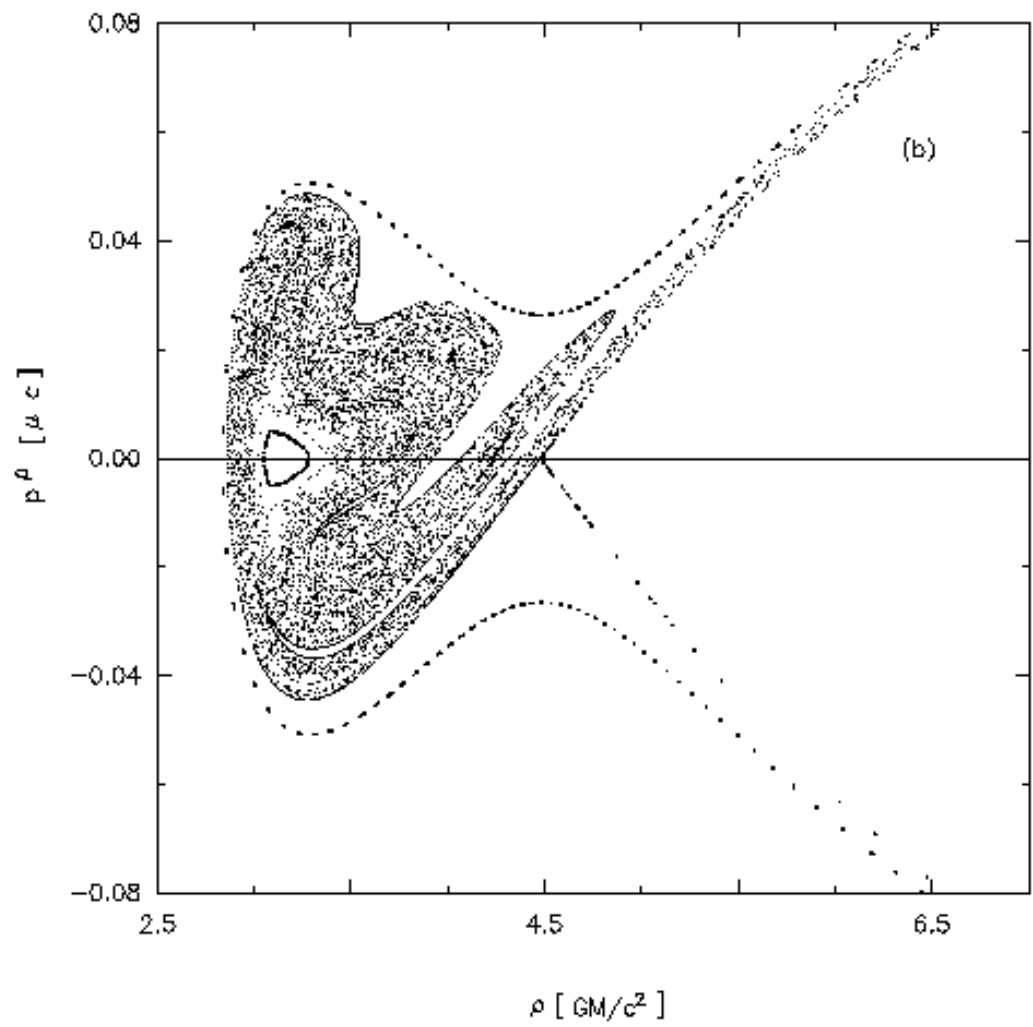
\end{document}